\documentclass{article}
\usepackage{spconf,amsmath,epsfig}
\usepackage{amsfonts}       
\usepackage{booktabs}

\renewcommand{\arraystretch}{1.2} 
\newcommand{\ra}[1]{\renewcommand{\arraystretch}{#1}}

\title{Scaling up Echo-State Networks with multiple light scattering}
\name{Jonathan Dong$^{1,2,4}$ \qquad Sylvain Gigan$^{1}$ \qquad Florent Krzakala$^{2}$ \qquad Gilles Wainrib$^{3}$
\sthanks{This research has received funding from the European Research Council under the EU’s 7th Framework Programme (FP/2007-2013/ERC Grant Agreement 307087-SPARCS and 278025-COMEDIA)}}
\address{$^{1}$ Laboratoire Kastler Brossel, CNRS UMR 8552, Ecole Normale Sup\'erieure, PSL Research University \\
Sorbonne Universit\'es \& Universit\'e Pierre et Marie Curie Paris 06, F-75005, Paris, France \\
$^{2}$ Laboratoire de Physique Statistique, CNRS, PSL Universités \& Ecole Normale Sup\'erieure, \\
Sorbonne Universit\'es et Universit\'e Pierre \& Marie Curie, 75005, Paris, France. \\
$^{3}$ Ecole Normale Sup\'erieure, D\'epartement d'Informatique, Paris, France. \\
$^{4}$ LightOn, 2 rue de la Bourse, 75002 Paris, France}

\begin{document}
%
\maketitle
\begin{abstract}
Echo-State Networks and Reservoir Computing have been studied for more than a decade. They provide a simpler yet powerful alternative to Recurrent Neural Networks, every internal weight is fixed and only the last linear layer is trained. They involve many multiplications by dense random matrices. Very large networks are difficult to obtain, as the complexity scales quadratically both in time and memory. Here, we present a novel optical implementation of Echo-State Networks using light-scattering media and a Digital Micromirror Device. As a proof of concept, binary networks have been successfully trained to predict the chaotic Mackey-Glass time series. This new method is fast, power efficient and easily scalable to very large networks.

\end{abstract}
\section{Introduction}

Since the work of Johnson and Lindenstrauss in 1984 \cite{johnson1984extensions}, random projections have been increasingly used in various settings. Their properties as low distortion transforms that eventually provide computational savings have made them useful in sketching high-dimensional datasets while speeding up various operations such as regression, clustering, embedding \cite{mahoney2011randomized,woodruff2014sketching}. A major issue with random projections stems in part due to the large number of operations needed to perform a multiplication between a random matrix and a feature vector. In order to remedy that bottleneck, a recent study has investigated the use of randomness found in natural physical processes to speed up these computations \cite{saade2015random}. Our present work explores the use of this randomness as a way to speed up an Echo State Network (ESN), a specific type of Recurrent Neural Network where neurons are connected with random weights \cite{jaeger2001echo,maass2002real,jaeger2004harnessing}. Complex dynamics are created in ESNs by the succession of random projections.

Recurrent Neural Networks are notoriously hard to train owing to the problem of vanishing or exploding gradients. To bypass this issue, researchers proposed Echo-State Networks where all the internal weights are fixed randomly, so that networks are a lot easier to train while we still keep the complex dynamics of Recurrent Neural Networks. The effectiveness of this strategy comes from the properties of random projections and this can be linked with recent works in Deep Learning where networks with fixed random weights are still able to perform well \cite{saxe2011random,rosenfeld2018}.

Initially inspired by neural networks, ESNs gave birth to Reservoir Computing (RC) \cite{lukovsevivcius2009reservoir}, a more general computational paradigm which has become popular in the last decade. An input sequence drives the complex dynamics of a reservoir that non-linearly encodes the input. The output is obtained by a linear combination of the reservoir state. To train such a system, one wants to find the best set of output coefficients, this step usually boils down to a linear regression. Such a framework has proven to be useful in tasks such as speech recognition, handwriting recognition, robot motor control or financial forecasting \cite{triefenbach2010phoneme,lukovsevivcius2012reservoir}. Furthermore, RC is not restricted to neural networks. Any physical dynamical system, even a bucket of water \cite{fernando2003pattern}, can be used for RC. Much work has been done to perform RC using non-linear optical elements \cite{vandoorne2008toward,woods2012optical,brunner2013parallel,vandoorne2014experimental,duport2016fully}, offering higher bandwidth and lower energy consumption.

Here we propose a novel hardware implementation of Reservoir Computing using a static complex medium -such as a layer of white paint- as a light-scattering material . Light is first modulated by a Digital Micromirror Device (DMD), then propagates through a multiply-scattering material, where it is subject to a large number of scattering events, the so-called multiple scattering regime. This simple experimental apparatus enables us to compute ESN states, as propagation through a complex medium can be modeled as a multiplication by a random matrix \cite{popoff2010measuring}. Like other optical implementations, this new approach can be very fast and only requires low power, as the multiplication by the random matrix, the most critical operation, is carried out "at the speed of light" and without power consumption. Lastly, the setup only uses off-the-shelf devices and a simple layer of scattering material. This simplicity makes it possible to replicate this approach in a lab experiment or in an integrated device.

Compared to previous optical implementations, it can potentially scale up the number of neurons very easily, the computational overhead being negligible. DMDs and cameras already routinely offer several million pixels of information. Hence, the number of neurons can potentially be increased up to these orders of magnitude. In this first demonstration, networks with 60,000 neurons are successfully trained for a speedup factor of 30 over high-end CPUs. Additionally, this optical implementation is closer to ESNs with a dense random weight matrix, in sharp contrast with implementations based on non-linear optical elements for which neuron connections are local. 

The DMD is a very fast programmable device that has many small tiltable mirrors with two orientations. Depending on their orientation, pixels are turned on (light sent on the diffusing material) or off (light sent on a beam blocker). As a consequence, this optical implementation will only use binary neurons. ESNs with binary neurons like in \cite{bertschinger2004real}, or binary ESNs, are commonly considered less powerful than real ESNs but good results can still be obtained thanks to the possibility of increasing the number of binary neurons, one advantage of this implementation. 

In this paper, we show the first implementation of a binary ESN using light-scattering materials. After a general presentation of ESNs in Section 2, Section 3 explains how to realize experimentally an optical ESN. In Section 4, we present and analyze the performance of this new implementation.


\section{Echo-State Networks}

We consider a network of $N$ binary neurons, writing the state $\mathbf{x}(t) = (x_1(t), \ldots, x_N(t)) \in \{-1; 1\}^N$ at timestep $t \in \mathbb{Z}$. Neuron $i$ receives an input from neuron $j$ with random weight $w_{ij} \in \mathbb{C}$ following a complex gaussian distribution. All neurons are interconnected and the system weights $\mathbf{W} = (w_{ij})$ is a dense random matrix. The input $\mathbf{i}(t)$ at time $t$ is also fed to every neuron with random weights $\mathbf{V}$. The output of a neuron is obtained by integrating the neuron inputs and applying a non-linear function $f$:
\begin{equation}
	\label{ESN iteration}
    \mathbf{x}(t+1) = f(\mathbf{W x}(t) + \mathbf{V i}(t))
\end{equation}
In the following, $f$ is a binary threshold on $|\mathbf{W x}(t) + \mathbf{V i}(t)|$ whereas ESNs traditionally used a hyperbolic tangent. The system is quite robust and other non-linearities are possible \cite{vandoorne2014experimental}. However, we will show that we need more binary networks to achieve a performance comparable to real networks. 

In a typical problem for ESNs, a time-dependent input $\mathbf{i}(t)$ of length $T$ and its corresponding output $\mathbf{o}(t)$ are given. During the training session, the reservoir is arbitrarily initialized and the time series $\mathbf{i}(t)$ is fed to the ESN. Driven by the input, the reservoir follows non-linear dynamics and the network states at every timestep are collected. The output predicted by the network at time $t$ is a linear combination of the state of the reservoir $\mathbf{W'x}(t)$ with output weights $\mathbf{W'}$ that are trained. Such a linear readout is very simple so that the training consists in a linear regression to find $\mathbf{W'}$ such that the error $\sum_t \| \mathbf{o}(t) - \mathbf{W'x}(t) \|^2$ is minimal. This convex optimization problem can be solved explicitly or iteratively. 

The dynamics of the reservoir needs to be finely tuned. Depending on the magnitude of the eigenvalues of the weight matrix, the reservoir can relax very quickly or follow a chaotic behavior. For maximal performance, hyperparameters need to be fixed to maximize the complexity of the ESN dynamics while keeping the stability of the system. Additionally, the first iterations are usually removed, as we wait for the network to forget the arbitrary initial state.

The computational bottleneck when running an ESN consists in the computation of the successive reservoir states. Eq. (\ref{ESN iteration}) involves a multiplication by the large random matrix $\mathbf{W}$ that needs to be repeated a large number of times. This prevents users from using a large number of neurons as the complexity scales quadratically with the neuron number, both in time and space. It is important to increase the reservoir size of an ESN because a larger reservoir can encode more subtle information on the input, and this leads to a better performance. In some implementations, sparse weight matrices can be used to speed up the calculation of this multiplication \cite{lukovsevivcius2009reservoir}. 



\section{Experimental realization}

When light propagates in a light-scattering medium, it does not go straight but scatters on inhomogeneities present at random and unpredictable positions. This complex propagation is traditionally viewed as an inconvenience that one wants to bypass, in order to image inside a thick biological sample for example. Here, we study and exploit this phenomenon with different point of view. Rather than trying to revert the changes caused by the complex medium, we want to make use of the complex image at the output called a speckle figure \cite{goodman1975statistical} to perform computation. Our interest is motivated by the speed and the scalability we can potentially obtain: we want to perform computations "at the speed of light". The large randomness of a speckle pattern has already been exploited for kernel methods \cite{saade2015random} or phase retrieval \cite{dremeau2015reference}.

\begin{figure}[h]
	\centering
	\includegraphics[width=.45\textwidth]{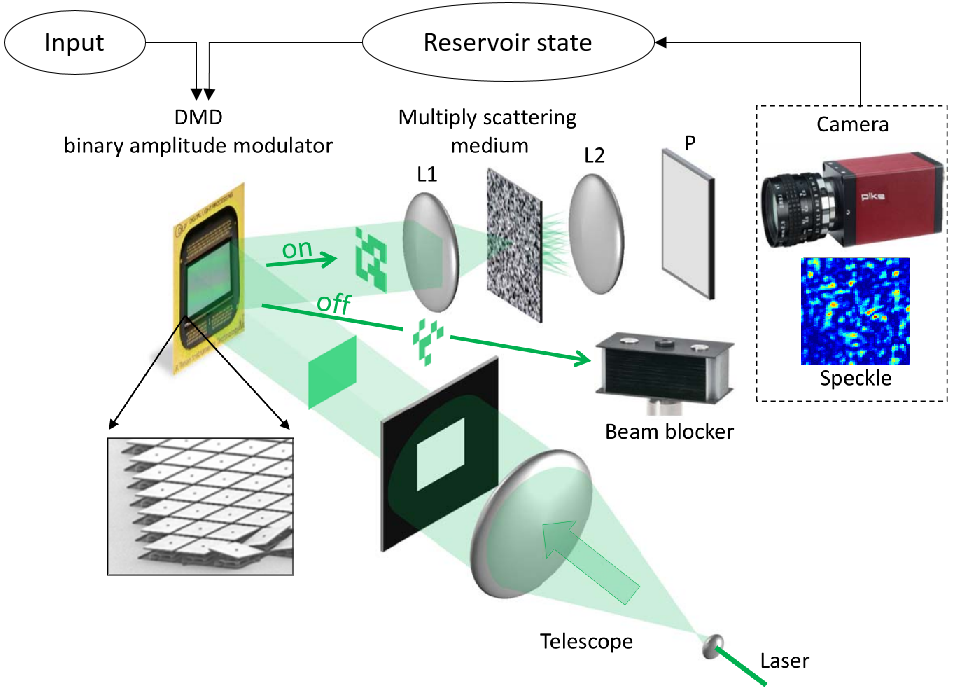}
	\caption{The experimental setup. Light from a laser is expanded by a telescope and arrives on a DMD. DMD pixels that are turned on send the coherent light on a multiply scattering medium. A speckle figure is collected on a camera and determines the next reservoir state. The reservoir state is then displayed with the input on the DMD to start a new iteration.}
	\label{Setup}
\end{figure}

Fig. \ref{Setup} shows the optical setup. A coherent light source from a continuous laser is expanded to fit on the DMD. The DMD is a programmable device that shapes the light and sends a binary image of size $M$ on the scattering medium. Here, the medium is a 2mm thick opaque layer, with $>10$ transport mean free path, where all light traversing the sample has been multiply scattered, and where absorption is negligible. The resulting speckle pattern is recorded by the camera.

This simple setup is useful to perform computation owing to the transfer matrix formalism. The electric field at the camera sensor $\mathbf{e} \in \mathbb{C}^N$ is a linear combination of the binary image $\mathbf{d} \in \{0;1\}^M$ sent by the DMD:
\begin{equation}
	\label{TM}
	\mathbf{e} = \mathbf{H} \mathbf{d}
\end{equation}
where $\mathbf{H}$ is the transfer matrix, an $N \times M$ random matrix. Thanks to the multiple scattering process, each element of $\mathbf{H}$ can be seen as an independent identically distributed random variable, drawn from a complex gaussian distribution, as demonstrated experimentally in \cite{popoff2010measuring} where this transfer matrix was measured. Thus, multiplication by a large random matrix is carried out by the scattering medium.

At every iteration, the current state of the ESN $\mathbf{x}(t)$ and the current input $\mathbf{i}(t)$ are displayed on the DMD. Necessarily binary, they are expanded and concatenated into a DMD image $\mathbf{d} \in \{0;1\}^M$ which is related to the electric field at the camera plane by (\ref{TM}). Cameras record a speckle intensity $\mathbf{s} \in \mathbb{R}^N$, equal to the modulus square of the electric field, $\mathbf{s} = |\mathbf{Hd}|^2 = |\mathbf{Wx}(t) + \mathbf{Vi}(t)|^2$, where $\mathbf{W}$ and $\mathbf{V}$ corresponds to subsets of columns of $\mathbf{H}$. 

From this intensity, the next ESN state $\mathbf{x}(t+1)$ is computed so that it satisfies (\ref{ESN iteration}). For this step, we consider a number of pixels of the camera image equal to the number of neurons. For every neuron, its new state is obtained after a threshold operation: the neuron is activated if the measured intensity of its corresponding pixel is greater than a threshold $A$ and silent otherwise. In other words, the activation function $f$ is a boolean function defined by $f(a) = (|a|^2 > A)$.

Once the next ESN state $\mathbf{x}(t+1)$ is obtained, a new iteration can start again. In a way, the speckle image determines what the DMD displays next. This whole process is repeated $T$ times where $T$ is the length of the input.


It is important to note that only two operations are computed optically. The scattering medium performs a multiplication by a random matrix and camera sensors record the modulus square of the electric field, i.e. they apply a non-linear operation. All the other operations like computing $\mathbf{x}(t)$ from $\mathbf{s}$ or the linear regression for training are performed on a computer.

This setup has first been built in a lab. Here we present results that were obtained using a high-performance implementation provided by LightOn and available on the cloud. This device is accessible for machine learning researchers to investigate problems that involve multiplications by dense random matrices.

\section{Results}

Echo-State Networks are designed for time-dependent tasks. We compare the performance of the optical setup and a CPU-based server on the Mackey-Glass time series prediction task. This time series is obtained by discretizing the following chaotic non-linear differential equation on $u$:

\begin{equation}
	\label{MG}
    \frac{du}{dt} = \beta \frac{u_\tau}{1+u_\tau^n} - \gamma u
\end{equation}

where $u_\tau$ represents the value of $u$ at time $t-\tau$. Here we take $\beta = 0.2$, $\gamma = 0.1$, $\tau = 17$, $n = 10$ and a discretization step $h = 1$. The input that is fed to the ESN is $i(t) = u(t)$ while the output corresponds to the prediction $o(t) = u(t+1)$.

The input needs to be displayed as a binary image on the DMD. Hence, we use a simple binary scheme to encode the real-valued input $u(t)$ in a binary vector of dimension 1000, in which the number of 1 is proportional to the value of $u(t)$. This binarization scheme increases the effective dimension of the input, but this does not affect the optical implementation. 

The camera records a speckle figure in \texttt{uint8} format, with values between 0 and 255. On this camera image, we choose a number of pixels equal to the ESN size, and we apply on these pixels a thresholding operation with threshold 24 to activate approximately half the neurons.

\begin{figure}[h]
	\centering
	\includegraphics[width=.45\textwidth]{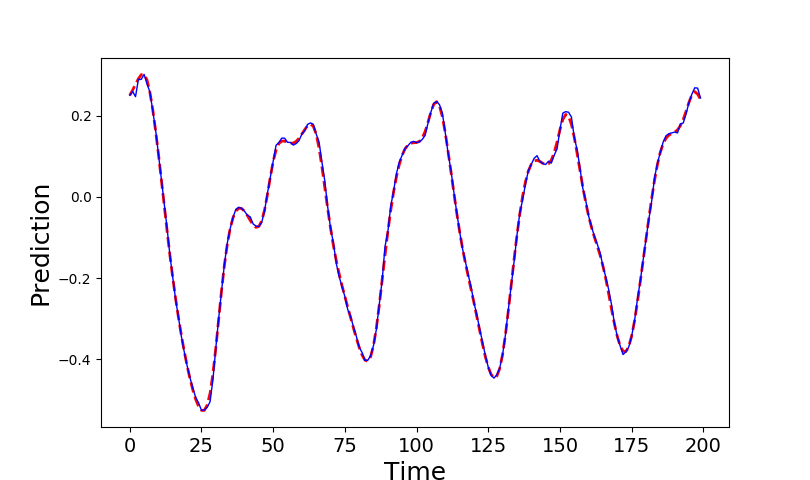}
	\caption{Prediction of a binary ESN of size 100'000 on Mackey-Glass dataset. Training has been performed on 2'000 timesteps with ridge regression parameter $\alpha$ set to 30.}
	\label{Prediction}
\end{figure}

\begin{table*}\centering
\ra{1.3}
\begin{tabular}{lllll}
\toprule
Method & ESN size & Init time & Time per 1000 iter & Performance \\
\midrule
CPU (Xeon E5-2690v3) & 47'315 & 234 & 720 & 0.995 \\
Optical (LightOn) & 10'000 & 0 & 3.2 & 0.971 \\
Optical (LightOn) & 100'000 & 0 & 3.2 & 0.985 \\
\bottomrule
\end{tabular}
\caption{Results of large-scale binary ESNs on Mackey-Glass dataset. Times are in seconds.}
\end{table*}

Optical computations are performed on a prototype provided by LightOn. This device is hosted in a datacenter, and gives remote access to a high-performance and stable implementation of the optical setup described here. Equivalent Echo-State Networks are also launched on a Microsoft Azure server with an Intel Xeon E5-2690v3 and 56 GB of RAM for comparison. The Mackey-Glass input is also binarized using the same scheme. This additional step has been introduced for comparison, and it has only a slight impact on the computational complexity for very large reservoir sizes. All the code is in Python using the scikit-learn syntax and training is done with the RidgeRegression class in scikit-learn. The performance corresponds to the score metric of regressors in scikit-learn. Higher is better: it measures the correlation of the prediction with the true output, the maximum is 1 for a perfect prediction.

To optimize the I/O bandwidth of the optical devices, we send images by batches of size between 300 and 3000. Between two batches, we collect the camera images on the computer and perform a thresholding operation to obtain the next image to be displayed on the DMD, which corresponds to the activation function of the ESN. This operation is very fast to compute compared to the multiplication by the weight matrix. Therefore, every batch corresponds to one iteration of a large number of ESNs in parallel and all the ESN states are concatenated for training. An even faster implementation is possible with direct communication between DMD and camera.

As seen in Fig. \ref{Prediction}, binary ESNs are able to learn how to predict the Mackey-Glass time series. On the Microsoft Azure server with 56 GB of RAM, the maximum size of an ESN is about 50'000. Above that number, there is not enough memory to keep the values of the random weight matrix. Moreover, at this size this dense random matrix takes 4 minutes to be created, due to the generation of a large amount of random numbers. On the other hand, with the optical implementation, there is no need to generate the weight matrix and we can go to much higher sizes. In the current setup, ESNs of size 100'000 are successfully trained optically. 

For ESNs of size 50'000, 2'000 iterations on the Microsoft Azure server take 24 minutes. With the same amount of time, we can perform 450'000 iterations optically (150 batches of 3000 images). Thus, the optical implementation is more than 225 times faster at this size. This number does not take into account the initialization of the random weight matrix, that is necessary without the optical setup. To compare performances, the linear regression for training is performed on 2'000 iterations. For very large reservoirs, this step starts to take more time and memory. 

\section{Conclusion}

This study presents a new physical implementation of ESNs, using light scattering to perform the fully connected random matrix multiplication, in very high dimensions. Non-linear time series predictions have been learned successfully by binary networks, both in experiments and simulations. 


This paves the way to ESNs that could be orders of magnitude faster and larger than feasible with silicon-only implementations. 

\section*{Acknowledgments}
We would like to thank LightOn (http://www.lighton.io/) for providing access to the optical hardware in the cloud. LightOn acknowledges support from OVH through the “OVH labs” program. FK ands SG acknowledges funding from the ERC under the European Union 7th Framework Program Grant Agreements 307087 and 724473.



\newpage

\bibliographystyle{IEEEbib}
\bibliography{Template}

\end{document}